 \documentclass[aps,pra,preprint,superscriptaddress,showpacs]{revtex4}

\usepackage{graphicx}
\usepackage{bm}

\begin{document}
\title{ Effect of $dt\mu$ quasi-nucleus structure on energy levels of the 
 $(dt\mu)Xee$ exotic molecule }

\author{O.~I.~Kartavtsev}
\affiliation{ Dzhelepov Laboratory of Nuclear Problems,
Joint Institute for Nuclear Research, Dubna, 141980, Russia }
\author{A.~V.~Malykh}
\affiliation{ Dzhelepov Laboratory of Nuclear Problems,
Joint Institute for Nuclear Research, Dubna, 141980, Russia }
\affiliation{ Physics Department, Novgorod State University, Novgorod 
the Great, 173003, Russia }
\author{V.~P.~Permyakov}
\affiliation{ Bogoliubov Laboratory of Theoretical Physics,
Joint Institute for Nuclear Research, Dubna, 141980, Russia }
\date{\today}

\begin{abstract}

Precise energies of rovibrational states of the exotic hydrogen-like molecule 
$(dt\mu )Xee$ are of importance for $dt\mu$ resonant formation, which is a key 
process in the muon-catalyzed fusion cycle. The effect of the internal 
structure and motion of the $dt\mu$ quasi-nucleus on energy levels is studied 
using the three-body description of the $(dt\mu)Xee$ molecule based on the 
hierarchy of scales and corresponding energies of its constituent subsystems. 
For a number of rovibrational states of $(dt\mu)dee$ and $(dt\mu)tee$, 
the shifts and splittings of energy levels are calculated in the second 
order of the perturbation theory. 

\end{abstract}

\maketitle

\section{Introduction}

It is known that one stopped muon in a deuterium-tritium mixture yields
more than 100 nuclear fusion reactions. The process of muon-catalyzed
fusion has been intensively studied and a detailed description exists
in the literature, e. g., in review 
articles~\cite{Breunlich89,Ponomarev90,Rafelski91,Froelich92}.
One of the key processes in the muon-catalyzed fusion cycle
is the formation of the hydrogen-like exotic molecule $(dt\mu )Xee$ 
(for the sake of generality $X$ stands for either isotope $d$, $t$, or $p$), 
in which a $dt\mu$ mesic molecule substitutes for one of the nuclei in the 
hydrogen molecule. It is widely accepted that the resonance mechanism 
proposed by Vesman~\cite{Vesman67} is responsible for the high rate of the 
$dt\mu$ formation. 
Due to this mechanism, a $dt\mu$ mesic molecule in a loosely bound excited 
state to be produced by low-energy collisions of $t\mu$ mesic atoms 
and ${\mathrm D}{\mathrm X}$  molecules in a resonance process
$$
t\mu + {\mathrm D} {\mathrm X} \longrightarrow (dt\mu)Xee 
$$ 
followed by $dt\mu$ transition to the $t\mu$ ground state.
The rate of the resonance process is sensitive to the
precise resonance position and an accuracy better a 1 meV is necessary
to obtain reasonable theoretical estimates of the formation rate 
\cite{Breunlich89,Ponomarev90,Rafelski91,Froelich92}. 

Resonance formation can take place if the energy released in $dt\mu$ 
binding is transferred to the rovibrational excitation of the exotic 
molecule $(dt\mu)Xee$. 
This is actually the case as $dt\mu$ has a loosely bound excited state with 
an angular momentum $\lambda = 1$ and binding energy which 
is comparable to vibrational quantum of the $(dt\mu)Xee$ molecule. 
In a non-relativistic approximation, different calculations determine 
with a good accuracy the binding energy of the isolated $dt\mu$ mesic 
molecule~\cite{Breunlich89,Ponomarev90,Rafelski91,Froelich92}. 
To obtain the precise value of the binding energy one has to correct 
the non-relativistic energy for relativistic effects, hyperfine effects, 
finite nuclear size, vacuum polarization, and others.
The resonance position is determined, besides the binding energy of 
isolated $dt\mu$, by the energy of the rovibrational excitation of 
the hydrogen-like molecule $(dt\mu)Xee$ with one nucleus being the particle 
$X$ and the other the excited $dt\mu$ mesic molecule. 
As the "size" of the excited $dt\mu$ mesic molecule with $\lambda = 1$ 
is of the order of 0.05 a.u.~\cite{Harston92}, which is much smaller than 
the internuclear distance in the whole molecule, the rovibrational spectrum 
of $(dt\mu)Xee$ can be calculated to a good approximation by treating 
$dt\mu$ as a point-like charged particle~\cite{Faifman86, Scrinzi88}. 
Nevertheless, to reach an accuracy of the order of a tenth of a meV one should 
take into account the energy shift which arises due to the internal structure 
and motion of a $dt\mu$ mesic molecule.  

The effect of the $dt\mu$ finite size was previously 
investigated in a simple approach~\cite{Menshikov85,Scrinzi89a,Harston92}
where the energy shifts for the $(dt\mu)dee$ were obtained by 
multiplying by 1.45 the shift calculated for the atom-like system $(dt\mu)e$
in the second order perturbation theory (PT). Within the framework of this 
simple approach it is not possible to take account of the molecular structure; 
in particular, the calculated energy shift is independent of the 
rovibrational quantum numbers. The effect of the molecular structure, i.~e., 
the dependence on angular momentum, was explicitly demonstrated 
in the elaborate six-body calculation~\cite{Harston92} of the 
$(dt\mu )dee$ energy shifts in the first order of the perturbation theory. 
Note, however, that the first- and second-order PT contributions to the 
energy shift are comparable. 
Recently, resonance positions in the low-energy $t\mu +{\mathrm D}_2$ 
scattering have been obtained in the elaborate three-body 
calculation~\cite{Zeman00,Zeman01}. Only few resonance states with the 
zero total angular momentum have been considered in this paper. 
 
The main aim of the present paper is to calculate the energy shifts 
which arise due to the internal structure and motion of the $dt\mu$ mesic 
molecule embedded in the hydrogen-like $(dt\mu)Xee$ molecule. 
The calculation is reduced to solution of a three-body problem for 
heavy particles $t\mu$, $d$ and $X$. 
This approach is based on the hierarchy of scales and corresponding energies 
of constituent subsystems of the $(dt\mu)Xee$ thus reliably taking into 
account the specific features of this molecule. 
As a result, the energy shifts are obtained for a number of vibrational 
and rotational states of $(dt\mu)dee$ and $(dt\mu)tee$ in the second-order PT.

\section{Method}


The structure of the exotic molecule $(dt\mu)Xee$ is characterized 
by a hierarchy of scales and corresponding energies of its constituent 
subsystems. In this respect, a $t\mu$ mesic atom is small in comparison with 
its mean separation from a deuteron in the loosely bound $dt\mu$ mesic 
molecule, which allows $t\mu$ to be treated as a point-like neutral particle  
interacting with a deuteron by the short-range effective potential. There is 
also interaction of $t\mu$ with the second nucleus $X$; however, this might be 
neglected due to large separation between these particles. 
In turn, the size of a $dt\mu$ mesic molecule is small in comparison with the 
amplitude of vibrations in $(dt\mu)Xee$; therefore, it moves as a point-like 
quasi-nucleus near the equilibrium position. For this reason, the effect 
of the $dt\mu$ structure is considered within the framework of the 
perturbation theory. 

Furthermore, two electrons in the hydrogen-like molecule $(dt\mu)Xee$ move 
much faster than the heavy particles $d$, $X$, and $t\mu$, which makes it 
possible to use 
the familiar Born-Oppenheimer (BO) approximation, i.~e., to solve 
electronic problem with the fixed charged particles $d$ and $X$ thus 
obtaining the BO energy which plays a role of the effective potential 
between $d$ and $X$. The electronic excitations, which require a considerable 
amount of energy~\cite{Sharp71}, are not taken into account for the low-energy 
processes under consideration. 

As a result, the description of $(dt\mu)Xee$ is reduced to solution of a 
three-body problem for three particles $t\mu$, $d$ and $X$. The interaction 
between charged $d$ and $X$ is described by the well-known BO potential for the
hydrogen molecule. In accord with the treatment of the $t\mu$ mesic atom 
as a point-like neutral particle, the present calculation does not explicitly 
use the $t\mu + d$ effective potential, rather the result is expressed via 
the low-energy $t\mu + d$ scattering phase shifts and characteristics of the 
$t\mu$ mesic molecule in the loosely bound excited state.

The $(dt\mu)Xee$ states are either true bound states or narrow resonances
if their energy is below or above the $t\mu + {\mathrm D}{\mathrm X} $ 
threshold.
 As the energy shifts are mainly determined by the coupling with closed 
channels, in the present calculation both resonances and bound states 
are treated on an equal footing thus neglecting a small contribution to 
the energy shifts which comes from the coupling with the open  
$t\mu +{\mathrm D}{\mathrm X}$ channel.

\subsection{Three-body description}

Under the above approximations, the Schr\"odinger equation 
for the hydrogen-like molecule $(dt\mu)Xee$ reads
\begin{equation}
\label{schredeq}
\left[ -\frac{1}{2\mu_1} \Delta_{r} - \frac{1}{2\mu_2}
\Delta_{\rho} + V_1( r) + V_2(|\bm{\rho} - \beta{\mathbf r}|) + V(|\bm{\rho} 
+ \alpha{\mathbf r}|) - E \right] \Psi = 0
\end{equation}
where the Jacobi coordinates ${\mathbf r}$ and $ \bm{\rho} $
are the vectors from $d$ to the point-like mesic atom $t\mu$ and from
the second nucleus $X$ to the $dt\mu$ center of mass, respectively.
The reduced masses and parameters $\alpha$ and $\beta$ are
$ \mu_1 = \displaystyle\frac{m_1m_2}{m_1+m_2}$,
$\mu_2 = \displaystyle\frac{(m_{1}+m_{2}) m_3}{m_1+m_2+m_3}$,
$\alpha = \displaystyle\frac{m_1}{m_1+m_2}$, and 
$\beta = \displaystyle\frac{m_2}{m_1+m_2}$, where $m_1, m_2$ and 
$ m_3$ are the masses of $t\mu$, $d$, and $X$, respectively. 
The atomic units are used throughout the paper unless other is specified.  
In Eq.~(\ref{schredeq}), $V(|\bm{\rho} + \alpha{\mathbf r}|)$ 
denotes the well-known BO potential describing the interaction between 
charged $d$ and $X$ while the short-range potentials $V_1(r)$ and 
$V_2(|\bm{\rho} - \beta{\mathbf r}|) $ describe the interaction 
of a $t\mu$ mesic atom with a deuteron and $X$, respectively. 
In the following, due to large internuclear separation ($\rho \gg r$) 
in $(dt\mu)Xee$, the short-range interaction 
$V_2(|\bm{\rho} - \beta{\mathbf r}|) $ of the $t\mu$ mesic atom 
with the second nucleus $X$ is negligible and will be omitted. 

A natural zeroth-order approximation for the calculation of the $(dt\mu)Xee$ 
energy levels is to treat the $dt\mu$ mesic molecule as a point quasi-nucleus 
with the $dt\mu$ mass and the unit charge. The calculations of the energy 
levels in this approximation are presented in~\cite{Faifman86, Scrinzi88} 
for different isotopes $X$ of the hydrogen-like molecule $(dt\mu)Xee$. 
Clearly, the treatment of $dt\mu$ as a point-like particle is equivalent 
to the replacement of the exact potential $V(|\bm{\rho} + \alpha{\mathbf r}|)$ 
in the Schr\"odinger equation~(\ref{schredeq}) by the potential $V(\rho) $ 
which describes the BO interaction between $X$ and the point particle located 
at the $dt\mu$ center of mass. Thus, the effect of the $dt\mu$ structure, 
which leads to the shift of the zeroth-order energy levels, originates from 
the perturbation potential 
\begin{equation}
\label{hamiltonian2}
V_{p} = V(|\bm{\rho} + \alpha{\mathbf r}|) - V(\rho) \ . 
\end{equation} 

In the zeroth-order approximation $V_{p} = 0$, the solutions of 
Eq.~(\ref{schredeq}) with the total angular momentum $L$ and its projection 
$M$ are written as a product of the bispherical harmonics 
${\cal Y}^{LM}_{l\lambda}(\hat{\bm{\rho}}, \hat{\bm{r}})$ describing 
the angular dependence, the radial function of $\rho$ describing the motion 
of nuclei in $(dt\mu)Xee$ with the angular momentum $l$, and the radial 
function of $r$ describing the internal motion in a mesic molecule with 
the angular momentum $\lambda$. The unperturbed energies $E_{nl}$ 
and the corresponding square integrable radial functions $\Phi_{nl}(\rho)$ 
of the $(dt\mu)Xee$ vibrational and rotational states satisfy the equation 
\begin{equation}
\label{radeqmol}
\left\{\frac{1}{2\mu_2}\left [-\frac{1}{\rho^2}\frac{\partial}{\partial \rho}
\left (\rho^2\frac{\partial}{\partial \rho}\right ) +
\frac{l(l+1)}{\rho^2}\right ] + V(\rho) - E_{nl}\right\}\Phi_{nl}(\rho) = 0 
\end{equation} 
where $n$ is the vibrational quantum number. For the problem under 
consideration, one should consider both the bound and continuum states 
of the $t+d\mu$ subsystem whose 
energies and wave functions satisfy the equation 
\begin{equation}
\label{radeqcont}
\left\{\frac{1}{2 \mu_{1}}\left [-\frac{1}{ r^2}\frac{\partial}{\partial r}
\left(r^2\frac{\partial}{\partial r}\right) +
\frac{\lambda (\lambda + 1)}{r^2}\right ] + V_1(r) - E\right\}\phi(r) = 0 \ . 
\end{equation} 
 Here  $E = -\varepsilon_{v \lambda}$ and 
$\phi (r) = \phi_{v \lambda}(r)$ for the bound states and 
$E = k^2/2\mu_1$ and $\phi (r) = \phi_{k \lambda}(r)$ for the continuum 
states with the wave number $k$. 
The functions $\phi_{v \lambda}(r)$ are square integrable and the functions 
$\phi_{k \lambda}(r)$ are normalized by the condition
\begin{equation}
\label{norm171}
\int\limits_{0}^{\infty}r^2 dr \phi_{k \lambda}^{\ast}(r) \phi_{q \lambda}(r)
= \delta(k-q) \ . 
\end{equation} 

In correspondence with the Vesman mechanism, 
 $(dt\mu)Xee$ contains a $dt\mu$ mesic molecule in the weakly bound 
state with the binding energy $\varepsilon_{11}$ $(v=1, \lambda=1)$. 
Other $dt\mu$ states, whose binding energies significantly exceed  
all the characteristic energies of the problem under consideration, 
will not be taken into account in the calculation of the energy shifts.

\subsection{Perturbation theory}

The effect of the $dt\mu$ structure is small due to smallness 
of $dt\mu$ mesic molecule in comparison with a characteristic length 
of $dt\mu$ motion in the molecular potential $V(\rho) $. In other words,  
the perturbation $ V_{p}$ is small in comparison with $V(\rho) $ and can be 
expanded in powers of the small parameter $\alpha r $. Correspondingly, 
the dimensionless parameter of the perturbation theory is the ratio 
of the average distance between the deuteron and the $dt\mu$ center of mass 
$\alpha \langle r \rangle$ to the average amplitude of vibrations 
$\langle \rho - a \rangle$ in the molecular potential near the equilibrium 
internuclear distance $a$. 

One should note that the lowest-order term of the expansion $V_p$, which is 
proportional to $\alpha r$, does not contribute to the energy shifts 
in the first-order PT; therefore, the energy shift of order $(\alpha r )^2$ 
must be obtained up to the second-order PT. Besides, $V_{p}$ couples the 
rotational states with $l = L \pm 1$ while the state with $l = L$ remains 
uncoupled. 
As the separation of the rotational levels is comparatively small, 
the level coupling cannot be {\em a priori} neglected and requires explicit 
treatment. Thus, the energy shifts will be determined in the second-order 
degenerate PT by solving a secular equation 
\begin{equation}
\label{matrixeq}
\det[V^{n} + W^{n} + {\cal E}_{n} - E] = 0
\end{equation}
where $V^{n}$ and $W^{n}$ are the matrices with the matrix elements 
of the first- and second-order PT $V^{n}_{ll_{1}}$ and $ W^{n}_{ll_{1}}$, 
respectively, the matrix elements of $ {\cal E}_{n}$ are 
$(E_{nl} + \varepsilon_{11}) \delta_{ll_{1}}$, and $E$ is the level energy. 

The first-order PT matrix elements are
\begin{equation}
\label{matrel1}
V_{ll_1}^{n } = \int d^{3}r d^{3}{\rho}  V_{p} \mid\phi_{11}(r)\mid^{2}
\Phi_{nl}(\rho)
\Phi_{n l_{1}}(\rho) {\cal Y}^{LM^{\ast}}_{l1}(\hat{\bm{\rho}}, \hat{\bm{r}})
{\cal Y}^{LM}_{l_11}(\hat{\bm{\rho}}, \hat{\bm{r}}) 
\end{equation} 
and the second-order PT matrix elements include a sum and an integral
over intermediate states describing simultaneous excitations 
of the exotic molecule with the quantum numbers $\nu$ and $\ell$ and a $dt\mu$ 
mesic molecule with the continuum-state wave number $k$ and the angular 
momentum $\lambda$ 
\begin{equation}
\label{matrel4}
W_{ll_1}^{n} = -\sum\limits_{\nu \ell \lambda }\int
\frac{dk Z_{n l,\nu \ell}^{\lambda}(k)
Z_{\nu \ell, n l_1}^{\lambda}(k)}
{k^2/2\mu_1 + E_{\nu \ell}-E} 
\end{equation} 
where \begin{equation}
\label{matrel2}
Z_{nl,n_1l_1}^{\lambda}(k) = \int d^{3}r d^{3}{\rho} V_{p} 
\phi_{11}(r) \phi_{k \lambda}(r) 
\Phi_{nl}(\rho) \Phi_{n_1 l_1}(\rho) 
{\cal Y}^{LM^{\ast}}_{l1}(\hat{\bm{\rho}}, \hat{\bm{r}}) 
{\cal Y}_{l_1\lambda}^{LM}(\hat{\bm{\rho}}, \hat{\bm{r}}) \ . 
\end{equation}

Bearing in mind that the second-order PT calculation are of order 
$\alpha^2 \langle r \rangle^2$, we expand $ V_{p}$ up to the second order 
in $\alpha r$, which corresponds to the multipole expansion 
\begin{equation}
\label{potVmod}
V_{p} = \alpha r \frac{\partial V}{\partial \rho} P_1(\cos\theta ) +
\frac{1}{6} \alpha^2 r^2
\left [\frac{\partial{^2}V}{\partial \rho^{2}} +
\frac{2}{\rho}\frac{\partial V}{\partial \rho} + 2\left (\frac{\partial^{2} V}
{\partial \rho^2}-\frac{1}{\rho}\frac{\partial V}{\partial \rho}\right )
P_2(\cos\theta )\right ]
\end{equation}
where $\theta $ is the angle between the vectors $\mathbf r $ and $\bm{\rho}$. 
The monopole and quadrupole terms in (\ref{potVmod}) contribute only 
in the first-order PT while the dipole one in the second-order PT. 

Calculation of the matrix elements $V_{ll_1}^{n}$ and $ W^{n}_{ll_{1}}$
with the perturbation $V_p$ in the form (\ref{potVmod}) results in
\begin{equation}
\label{matrel1mod}
V_{ll_1}^{n} = \frac{1}{6}\alpha^{2}Q[U^{M}_{nl,n_1l_1}\delta_{ll_1} + 
U^{Q}_{nl,n_1l_1} A_{2}^{L}(l 1 l_1 1)] \ , 
\end{equation}  
\begin{equation}
\label{W} 
W_{ll_1}^{n} = -\alpha^2\sum_{\nu}U_{n\nu}^{D}U_{\nu n}^{D}\sum_{\lambda \ell}
A_1^L(l1\ell\lambda)A_1^L(l_11 \ell\lambda )I_{\lambda}(E_{\nu \ell} - E + 
\varepsilon_{11}) 
\end{equation}
where
\begin{equation}
\label{Idelta}
I_{\lambda}(\Delta) = \int\limits_{0}^{\infty}
 \frac {\left[u_{\lambda}(k)\right]^2dk}
{k^{2}/2\mu_1+\Delta} \ , 
\end{equation}
\begin{equation}
\label{Q}
Q = \int dr r^4 \mid \varphi_{11}(r)\mid^{2} \ ,
\end{equation}
\begin{equation}
\label{vm}
u_{\lambda}(k) = \int dr r^{3} \phi_{k \lambda}(r) \varphi_{11}(r) \ ,
\end{equation}
\begin{equation}
\label{U2}
U^{M}_{nl,n_1l_1} = \int \rho^2d\rho \Phi_{nl}(\rho) \Phi_{n_1 l_1}(\rho)
\left(\frac{\partial^{2} V}{\partial \rho^2}+\frac{2}{\rho}
\frac{\partial V}{\partial \rho}\right) \ ,
\end{equation}
\begin{equation}
\label{U3}
U^{Q}_{nl,n_1l_1} = 2\int \rho^2 d\rho \Phi_{nl}(\rho) \Phi_{n_1 l_1}(\rho)
\left(\frac{\partial^{2} V}{\partial \rho^2} - 
\frac{1}{\rho}\frac{\partial V}{\partial \rho}\right) \ ,
\end{equation}
\begin{equation}
\label{U1}
U^{D}_{nl,n_1l_1} = \int\rho^2 d\rho \Phi_{nl}(\rho)
\Phi_{n_1l_1}(\rho) \frac{\partial V}
{\partial \rho} \ .
\end{equation} 
and the angular integrals $A_{K}^{L}(l\lambda l_1 \lambda_{1})$
are given in the Appendix. 

\section{Results of calculation}

\subsection{Matrix elements}

The simple, though providing the required accuracy expressions 
for the multipole matrix elements~(\ref{U2}), (\ref{U3}), and (\ref{U1}) 
are obtained using the following reliable approximations. Firstly, 
the matrix elements are completely determined by the BO potential 
for the hydrogen molecule $V(\rho)$ which is fairly well known from the 
calculations~\cite{Kolos64, Kolos86, Faifman86, Scrinzi88}. As 
$(dt\mu)Xee$ is produced in low-energy $t\mu + \mathrm{D X}$ collisions, 
only the lowest vibrational states should be taken into account. For these 
states are localized near the minimum of $V(\rho)$ at the equilibrium 
internuclear distance $a \approx 1.4$~a.u., it is natural to use the harmonic 
approximation 
\begin{equation}
\label{Vaprox1}
V_h(\rho) = \frac{1}{2} \mu_2 \omega^2 (\rho - a)^2 + V_{0}
\end{equation} 
where only the frequency of vibrations $\omega $ is of importance for 
the calculation. Besides, an accuracy of the approximation~(\ref{Vaprox1}) 
is estimated using the unharmonic approximation of the BO potential 
\begin{equation}
\label{Vaprox2}
V_u(\rho) = \frac{1}{2} \mu_2 \omega^2 (\rho - a)^2 [1 - \alpha_M (\rho - a)]
+ V_{0} \ .
\end{equation} 
which takes into account the next term of the expansion in $\rho - a$. 
The approximation (\ref{Vaprox2}) accurately reproduces the exact energies 
of the lowest vibrational states calculated in~\cite{Faifman86, Scrinzi88}.

Secondly, the rotational energy in (\ref{radeqmol}) for the hydrogen-like 
molecule $l(l+1)/2\mu_{2}\rho^{2} \approx l(l+1)/2\mu_{2}a^{2} 
\approx 10^{-4}$ is two orders of magnitude smaller than the vibrational 
energy $\omega \approx 10^{-2}$. Therefore, under a usual approximation, 
the centrifugal term is treated perturbatively, i.~e., the eigenenergies 
are given by 
\begin{equation}
\label{Eaprox}
E_{nl} = E_{n0} + v_{r}l(l+1) 
\end{equation} 
and the wave functions $\Phi_{nl}(\rho)$ will be taken independent of $l$ 
in the same approximation. Indeed, the rotational spectrum calculated in 
\cite{Faifman86, Scrinzi88} is in good agreement with the above 
expression~(\ref{Eaprox}) with 
$v_{r} \approx 1/2\mu_{2}a^{2} \simeq 10^{-4}$. Thus, under the above 
approximations, the radial wave function $\Phi_{nl}(\rho)$ 
in the potential~(\ref{Vaprox1}) coincides with the harmonic-oscillator 
wave function and the multipole matrix elements (\ref{U2}), (\ref{U3}), 
(\ref{U1}) are reduced to $l$-independent expressions 
\begin{equation}
\label{U100}
U_{n \nu}^{D} = \sqrt{\frac{\mu_2  \omega^3}{2}}\left( \sqrt{n}
\delta_{n-1,\nu} + \sqrt{n+1}\delta_{n+1,\nu}\right), \qquad U_{nn}^{M} = 
\mu_2 \omega^2, \qquad U_{nn}^{Q} = 2 \mu_2 \omega^2 \ . 
\end{equation} 
The unharmonic term of the potential $V_u(\rho)$ leads only to modification of 
the dipole matrix element 
\begin{eqnarray}
\label{U1an}
& & U_{n\nu}^{D} = \sqrt{\frac{ \mu_{2} \omega^{3}}{2}}
\biggl\{ \left[\sqrt{n}
\delta_{n-1,\nu} + \sqrt{n+1}\delta_{n+1,\nu}\right] - \nonumber\\
& - & \eta\left [\sqrt{n(n-1)}\delta_{n,\nu+2}
+\sqrt{(n+1)(n+2)} \delta_{n,\nu-2}+(2n+1)\delta_{n,\nu} \right ] \biggl\}
\end{eqnarray} 
where the unharmonic correction is proportional to the dimensionless 
parameter
$\eta = \frac{3}{2}\alpha_{M}/ \sqrt{2 \mu_{2}\omega} \approx 0.14$.

Calculation of the quasi-nucleus matrix elements~(\ref{Q}, \ref{vm}) is based 
on the smallness of the $t\mu$ size in comparison with the size of the loosely 
bound $dt\mu$ state ($v = 1, \lambda = 1$). Thus, almost in all 
the configuration space $t\mu$ and $d$ move as free particles and 
the bound-state wave function is approximated by 
\begin{equation}
\label{asimptfun}
\varphi_{11}(r) = C_{a} \frac{1 + \kappa r}{\sqrt{\kappa} r^{2}} e^{-\kappa r} 
\end{equation}
where $\kappa = \sqrt{2\mu_{1}\varepsilon_{11}}$. The asymptotic 
expression~(\ref{asimptfun}) has been widely used in description of 
the loosely bound states of mesic molecules and the asymptotic 
normalization constant $C_{a}$ was determined by a comparison with the exact
three-body calculations~\cite{Menshikov85,Aissing90,Kino95}. 
For the same reasons, the asymptotic expressions are used
for the continuum wave functions $\phi_{k \lambda}(r)$, viz., 
\begin{equation}
\label{asimptcont}
\label{asimptcont0}
\phi_{k 0}(r) = \sqrt{\frac{2}{\pi}} \frac{\sin( k r+\delta_{0}(k))}{r} \ , 
\end{equation}
\begin{eqnarray}
\label{asimptfun2}
\phi_{k2}(r) = \sqrt{\frac{2}{\pi}}k\left[\cos\delta_2(k) j_{2}(kr) +
\sin\delta_{2}(k) y_{2}(kr)\right]
\end{eqnarray}
where $\delta_{\lambda }(k)$ are the $t\mu + d$ scattering phase shifts 
and $ j_{2}(kr)$ and $y_{2}(kr)$ are the spherical Bessel 
functions. The $d$-wave phase shift $\delta_{2}(k)$ is actually very small, 
which allows either replacing $y_{2}(kr)$ by the leading term 
$\displaystyle\frac{\cos kr}{kr}$ of its asymptotic expansion or simply 
putting $\delta_{2}(k) = 0$ in Eq.~(\ref{asimptfun2}). Using the wave 
functions (\ref{asimptfun}), (\ref{asimptcont}), and (\ref{asimptfun2}) with 
$y_{2}(kr) \to \displaystyle\frac{\cos kr}{kr}$ one obtains 
the quadrupole momentum 
\begin{equation}
\label{Qasimpt}
Q = \frac{5}{8} \frac{C_a^2}{\mu_{1}\varepsilon_{11}}
\end{equation}\\
and the expression 
\begin{equation}
\label{J5L}
I_{\lambda}(\Delta) = \frac{4C_{a}^{2}}{\pi \mu_{1}\varepsilon_{11}^{2}}
J_{\lambda}(\Delta/\varepsilon_{11})
\end{equation}
via the dimensionless integrals 
\begin{eqnarray}
\label{J0L}
J_{0}(z) &=& \int\limits_{0}^{\infty} \frac{\left [ \sin\delta_0(k) - 
(k^2+3)(k/2)\cos\delta_0(k) \right ]^2dk}
{(k^2+1)^4(k^2+z)}\ ,\\ 
\label{J2L}
J_{2}(z) &=& \int\limits_{0}^{\infty} \frac{\left[ \sin\delta_2(k) + 
k^3\cos\delta_2(k) \right]^2dk}
{(k^2+1)^4(k^2+z)}\ .
\end{eqnarray}

\subsection{Shift and splitting of energy levels }
\label{secshifts}

Energy shifts are obtained by solving the secular equation~(\ref{matrixeq}) 
which is reduced, due to the selection rules for angular momenta $l$ 
and $l_{1}$, to a $2\times2$ matrix equation for $l, l_1 = L\pm1$ ($L \ne 0$) 
and a scalar equation for $l = l_1 = L \ne 0$. The energy shifts with 
respect to the unperturbed rovibrational energies $E_{nl} + \varepsilon_{11}$ 
are denoted as $\Delta_{0}(n)$ and $\Delta_{\pm}(nl)$ for $l = L$ and 
$l = L \pm 1$, respectively. Note that the state with $L = 0$ and 
$l = l_1 = 1$ is uncoupled; however, its energy shift $\Delta_{+}(n1)$ will 
be determined in the same manner as for the other $L \ne 0$ states. 

The first-order PT matrix elements $V^{n}_{ll_{1}}$ in Eq.~(\ref{matrixeq}) are
calculated by substituting the radial integrals $U^{M,Q}_{nn}$~(\ref{U100}), 
the quadrupole momentum $Q$~(\ref{Qasimpt}), and the angular integrals 
$A^{L}_{2}(l1l_{1}1)$~(\ref{Aquadro}) in Eq.~(\ref{matrel1mod}). Note that 
$V_{ll_1}^{n}$ appears to be independent of the vibrational quantum number 
$n$ and this index will be omitted in what follows. 
The matrix elements are scaled by a single dimensional parameter
\begin{equation}
\label{v_0}
v_0 = \frac{m_{1} m_{3} \omega^2 C_{a}^2}{16m_{2} (m_1+m_2+m_3) 
\varepsilon_{11}}
\end{equation}
which is a characteristic energy for the problem. It should be mentioned that 
although Eq.~(\ref{v_0}) does not contain a specific small parameter, 
$v_0$ turns out to be sufficiently small $(v_0/\omega \sim 0.006)$ thus making 
the energy shifts small. As a result, one finds 
\begin{equation}
\label{Vll}
V_{ll_{1}} = v_0 \left\{
\begin{array}{ll}
1, & \quad l, l_1 = L \quad (L \neq 0) \\ 
2\delta_{ll_{1}} - B_{ll_{1}}, & \quad  l,l_{1} = L\mp 1  
\end{array} 
\right. 
\end{equation}
where the matrix elements $B_{ll_{1}}$ form the matrix 
\begin{equation}
\label{V600}
B = \frac{1}{2L+1}
\left(\begin{array}{cc}1 & 2\sqrt{L(L+1)}\\
2\sqrt{L(L+1)} & -1\end{array}\right) 
\end{equation} 
in which the first row and column correspond to $l, l_1 = L - 1$ 
and the second ones to $l, l_1 = L + 1$.

The second-order PT matrix elements $W_{ll_{1}}^{n}$ in Eq.~(\ref{matrixeq}) 
are calculated by substituting $U_{n \nu}^{D}$~(\ref{U100}) and 
$I_{\lambda}(\Delta)$~(\ref{J5L}) in Eq.~(\ref{W}), which gives the expression 
\begin{equation}
\label{WN}
W_{ll_1}^{n} = -v_{0} \frac{32\omega}{\pi\varepsilon_{11}}\sum_{\nu=n\pm1}
\sum_{\ell \lambda} \max(n,\nu) A_1^L(l1 \ell \lambda)
A_1^L(l_1 1 \ell \lambda )J_{\lambda}
\left(1 + \frac{E_{\nu \ell} - E}{\varepsilon_{11}}\right)
\end{equation} 
via the energy scale $v_{0}$ and the dimensionless factors. 
Solving the secular equation~(\ref{matrixeq}), one can safely replace, 
up to an accuracy of the second-order PT, the eigenvalue $E$ in the argument 
of $J_{\lambda}$ by the unperturbed value $E_{nl}$. Thus, the calculation 
of the energy shifts is basically accomplished by derivation 
of Eqs.~(\ref{Vll}-\ref{WN}). 

However, it is reasonable to make further simplification of (\ref{WN}) 
by neglecting the difference of the rotational energies in the argument 
of $J_{\lambda}$, which allows obtaining an explicit and sufficiently accurate 
dependence of the energy shifts on the quantum numbers $n$ and $l$. 
As the rotational energy is much smaller than the vibrational quantum 
$\omega$, one replaces the energy differences $E_{n \pm 1 l} - E$ 
in the argument of $J_{\lambda}$ by the $l$-independent values 
$E_{n \pm 1 0} - E_{n 0} = \pm \omega$. Using the angular integrals 
$A^{L}_{1}(l1l_{1}\lambda)$~(\ref{Ad0}, \ref{Ad2}) and introducing the notation
$J^{\pm}_{\lambda} = J_{\lambda}\left(1 \pm \omega /\varepsilon_{11}\right) $ 
for integrals independent of $n$ and $l$ one obtains 
\begin{equation}
\label{Wll}
W_{ll_1}^{n} = v_0 \left\{
\begin{array}{ll}
1+\alpha_{n}-\beta_{n}, &\quad l,l_{1} = L \quad (L \neq 0) \\ 
(2-\beta_{n})\delta_{ll_1} + 
(\alpha_{n}-1)B_{ll_1}, &\quad  l,l_{1} = L \mp 1  
\end{array} 
\right. 
\end{equation}
where
\begin{eqnarray}
\label{albeta}
\alpha_{n} = 1 - \frac{16\omega}{3\pi \varepsilon_{11}}
\left[(n+1)(J_{0}^{+}+\frac{1}{5}J_{2}^{+}) + 
n(J_{0}^{-} + \frac{1}{5}J_{2}^{-})\right] \ , \\ 
\label{albeta1}
\beta_{n} = 2 - \frac{16\omega}{3\pi \varepsilon_{11}} \left [(n+1)(J_{0}^{+}
+ \frac{7}{5}J_{2}^{+}) + n(J_{0}^{-} + \frac{7}{5}J_{2}^{-})\right ] 
\end{eqnarray} 
determine the explicit dependence on the vibrational quantum number $n$.
As a result, the sum of $V_{ll_1}$~(\ref{Vll}) and $W_{ll_1}^{n}$~(\ref{Wll}) 
takes a simple form
\begin{equation}
\label{Ful}
V_{ll_1} + W_{ll_1}^{n} = v_0 \left\{
\begin{array}{ll}
\beta_{n} - \alpha_{n}, &\quad l = l_1 = L \quad (L \neq 0) \\ 
\beta_{n}\delta_{ll_1} - \alpha_{n}B_{ll_1}, &\quad  l,l_{1} = L\mp 1  
\end{array} 
\right. \ ,
\end{equation}
i.~e., the parameter $\beta_{n}$ determines the constant shift $v_{0}\beta_{n}$
of all level energies $E_{nl}$
whereas $\alpha_{n}$ determines the level splitting. Using (\ref{Ful})  and 
(\ref{Eaprox}) in the secular equation~(\ref{matrixeq}) one obtains 
\begin{eqnarray}
\label{scalar}
\Delta_{0}(n)& = &v_{0}(\beta_{n}- \alpha_{n}) \ , \\ 
\label{matr}
\Delta_{\pm}(nl)& = &v_0 \beta_{n} \mp v_r[2(l \mp 1)+1] \pm
\sqrt{v_{0}^{2}\alpha_{n}^2 +2v_{0}\alpha_{n}v_{r}+
[2(l \mp 1)+1]^{2}v_{r}^2} \ .
\end{eqnarray}
The effect of coupling of the rotational states with $l = L \pm 1$ 
is explicitly taken into account in expression~(\ref{matr}). 
Generally, the effect decreases with decreasing ratio of the level 
splitting to the energy difference between the rotational states 
$\displaystyle{v_0 \alpha_n \over v_r (2L + 1)}$, i.~e., with increasing
total angular momentum $L$. As follows from the numerical values of 
$v_{0}$, $v_{r}$, and $\alpha_{n}$ for all the considered states $n=2,3$ 
(Section~\ref{Numresults}), even in the worst case $L = 1$ the 
energy shifts calculated with and without allowance for the coupling of 
the rotational states differ at the most by $0.01$meV for $(dt\mu)dee$ 
and $0.03$meV 
for $(dt\mu)tee$. As these values are beyond the accuracy of the present 
calculation, it is quite reasonable to neglect coupling, i.~e., to use 
the diagonal approximation for the secular equation~(\ref{matrixeq}), 
which allows obtaining a simple expression 
\begin{eqnarray}
\label{enshift}
\Delta_{\pm}(nl) =
v_{0}\left[\beta_{n} \pm \frac{\alpha_{n}}{2(l \mp 1) + 1}\right] = 
v_{0}\left[\beta_{n} \pm \frac{\alpha_{n}}{2L + 1}\right] \ . 
\end{eqnarray}  

Note that Eqs.~(\ref{matr}) and~(\ref{enshift}) are valid both for $L = 0$ 
and $l = 1$ when $\Delta_{+}(n1) = v_{0}(\beta_{n} + \alpha_{n}) $ and for 
$L = 1$ and $l = 0$ when $\Delta_{-}(n0) = v_{0}(\beta_{n} - \alpha_{n}/3)$. 
The sign of $\alpha_{n}$ determines the relative position of the levels so 
that the energies satisfy the inequalities $E_{nL+1} > E_{nL-1} >E_{nL}$ for 
$\alpha_{n} > 0$ and the inverse inequalities for $\alpha_{n} <0$. The largest 
energy splitting is predicted for $l = 1$ between the states with $L = 0$ and 
$L = 1$, viz., $\Delta_{+}(n1) - \Delta_{0}(n) = 2 v_0 \alpha_{n}$. 

\subsection{Numerical results }
\label{Numresults} 

The energy shifts and level splittings will be calculated by solving 
the eigenvalue equation~(\ref{matrixeq}) using formulas~(\ref{v_0} 
- \ref{WN}). In addition to the particle masses 
$m_\mu = 206.768$a.u., $m_d = 3670.484$a.u., and $m_t = 5497.922$a.u.,
calculation of the matrix elements $V_{ll_1}$ and $W_{ll_1}^{n}$ requires  
the vibrational $\omega$ and rotational $v_r$ energies of the exotic molecule 
$(dt\mu)Xee$, the binding energy $\varepsilon_{11}$ and the asymptotic 
constant $C_a$ of the $dt\mu$ loosely bound state, and the low-energy 
$t\mu + d$ scattering phase shifts $\delta_\lambda (k)$ which determine the 
integrals 
$J_{\lambda } \left(1+\frac{E_{\nu \ell}-E}{\varepsilon_{11}}\right)$. 

The vibrational quantum $\omega $ and the rotational-energy constant $v_r$
are determined by the BO internuclear potential of the hydrogen
molecule near its minimum or, equivalently, by the low-lying part
of the $(dt\mu)Xee$ vibrational-rotational spectra calculated
in~\cite{Faifman86,Scrinzi88}.
Fitting the BO potential near the equilibrium distance $a = 1.401$ 
to the harmonic, unharmonic, and Morse potentials provides consistent
determination of both $\omega$ and the parameter $\alpha_M $.
As the BO potential is independent of the isotopic composition, both 
$\mu_2 \omega^2$ and $\alpha_M$ are independent of the masses of heavy 
particles due to Eq.~(\ref{Vaprox2}). The result of the fit gives 
$\omega = 321.8$meV for $(dt\mu)dee$ (correspondingly, $\omega = 273.1$meV for 
$(dt\mu)tee$) with a few per cent accuracy and the parameter 
$\alpha_M = 0.7$. For these parameters, the energies of the lowest 
vibrational states in the approximate potential are in reasonable agreement
with the results of~\cite{Faifman86,Scrinzi88}. 

The rotational spectra calculated in~\cite{Faifman86,Scrinzi88} 
are fitted to Eq.~(\ref{Eaprox}) for $1 \le l \le 10 $ and each 
$1 \le n \le 4 $. For the lowest vibrational state $n = 1$, one obtains
$v_r = 2.43$meV for $(dt\mu)dee$ and $v_r = 1.85$meV for $(dt\mu)tee$. 
These values agree with the simple estimate $2 \mu_2 v_r \approx 1/a^2$ that 
determines the isotopic dependence of $v_r$. Although $v_r$ slightly decreases 
for the higher vibrational states, the above values will be used for $n > 1$, 
which leads to a few per cent error. 

Determination of the binding energy $\varepsilon_{11}$ of a $dt\mu$
loosely bound state was a subject of numerous elaborated calculations.
As a result, the value $\varepsilon_{11} = 596$meV
\cite{Breunlich89,Ponomarev90,Froelich92} is obtained
for the lowest hyperfine state by taking into account relativistic effects,
hyperfine effects, finite nuclear size, and vacuum polarization.
The asymptotic constant $C_a$ was determined in
a number of papers~\cite{Menshikov85,Aissing90,Kino95} by a comparison of
the asymptotic expression~(\ref{asimptfun}) with the three-body
wave function. In the following, it is accepted the value 
$C_a = 0.874/\sqrt{2}$
obtained in the latest elaborated calculation~\cite{Kino95} of the wave 
function in a wide asymptotic region of large distances between $d$ and 
$t \mu $.
Using $C_a$,  $\varepsilon_{11}$, and $\omega$ one can calculate 
the energy scale $v_0$~(\ref{v_0}). As $\omega^2 \sim 1/\mu_2 =
\displaystyle{m_1 + m_2 + m_3 \over (m_{1} + m_{2}) m_3}$, 
the parameter $v_0$~(\ref{v_0}) is independent of $m_3$, i.~e., it is the same
for any isotope $X$. Given the above numerical values one obtains 
$v_0 = 1.81$meV. 

For the sake of completeness, it is interesting to estimate the energy scale
for the molecule $(dd\mu )Xee$ too by using the values
$\varepsilon_{11} = 1975$meV, $C_a = 1.006/\sqrt{2}$, and $\omega = 257$meV,
which gives $v_0 = 0.4$meV. Although the present approach requires some 
modifications to describe $(dd\mu )Xee$, viz., taking into account the 
identity of nuclei in $dd\mu $ and the essential role of the unharmonic 
corrections to the BO potential, one can qualitatively conclude that the 
energy shifts in $(dd\mu )Xee$ are $4-5$ times smaller than in $(dt\mu )Xee$. 

In the present approach the energy shifts in the first-order
PT are given by simple dependence on the angular momentum $l$
(\ref{Vll} - \ref{V600}) containing a single parameter $v_0$. It is worthwhile 
to compare this result with the first ever elaborate six-body calculation 
of the $(dt\mu )dee$ energy shifts in the first-order PT~\cite{Harston92}. 
In this paper, the molecular structure, i.~e., the dependence on $l$,
was explicitly taken into account in contrast with previous 
calculations~\cite{Menshikov85,Scrinzi89a,Harston92a} where the $l$-independent
energy shift was obtained by scaling the result for the atom-like four-body 
system $(dt\mu)e$. As pointed out in this paper, the monopole contribution 
calculated in Ref.~\cite{Harston92} depends on the choice 
of the coordinate system that does not allow a comparison. For this reason, 
only the quadrupole contribution to the first-order PT energy shifts of 
Ref.~\cite{Harston92} will be compared with the present results. 
The quadrupole contribution of Ref.~\cite{Harston92} is given in Table II 
of that paper, while in the present approach 
$\Delta E_{Q}^{(1)} = (10 / 3) v_0 A_{2}^{L}(l1l1)$, as follows from 
Eqs.~(\ref{matrel1mod}), (\ref{U100}), and (\ref{Qasimpt}). The both results 
are in excellent agreement with each other, as shown in Table~\ref{tab1}. 
\begin{table}[hbt]
\begin{tabular}{r r r r r c}
$l$ & \phantom{a} $L$ & $A_{2}^{L}(l1l1)$ & $\Delta E_{Q}^{(1)} $ &
$\Delta E_{Q}^{(1)}$\protect\cite{Harston92} & $\tilde v_{0} $  \\
\hline
$1$ & $0$ & $2/5$ \phantom{a.}&$2.42$\phantom{.} & $2.35$\phantom{a.} &$1.77$\\
$1$ & $1$ & $-1/5$\phantom{a.}&$-1.21$\phantom{.}& $-1.17$\phantom{a.}&$1.76$\\
$2$ & $1$ & $1/5$ \phantom{a.}&$1.21$\phantom{.} & $1.17$\phantom{a.} &$1.76$\\
$1$ & $2$ & $1/25$\phantom{a.}&$0.24$\phantom{.} & $0.23$\phantom{a.} &$1.73$\\
$2$ & $2$ & $-1/5$\phantom{a.}&$-1.21$\phantom{.}& $-1.17$\phantom{a.}&$1.76$\\
$3$ & $2$ & $4/25$\phantom{a.}&$0.96$\phantom{.} & $0.94$\phantom{a.} &$1.77$\\
\hline
\end{tabular}
\caption{Quadrupole contributions $\Delta E_{Q}^{(1)}$ (in meV)
to the first-order PT energy shifts of the present calculation and those
from Ref.~\cite{Harston92} for different $l$ and $L$. Also presented are
the angular integrals $A_{2}^{L}(l1l1)$ and the parameter $\tilde v_{0} $
corresponding to the energy shifts of Ref.~\cite{Harston92}. }
\label{tab1}
\end{table} 
Note that in the present approach the dependence on angular momenta 
is completely determined by the factor $A_{2}^{L}(l1l1) $ which is also 
presented in Table~\ref{tab1}. To a good accuracy, the results 
of Ref.~\cite{Harston92} reveal the same dependence on angular momenta 
which approves the description of energy shifts by a single parameter $v_{0}$. 
To emphasize this fact, the quadrupole correction calculated 
in Ref.~\cite{Harston92} is expressed in the form 
$\Delta E_{Q}^{(1)} = (10 / 3) \tilde v_0 A_{2}^{L}(l1l1)$ with the variable 
$\tilde v_{0} $ presented in Table~\ref{tab1}. Indeed, $\tilde v_{0} $ 
is practically independent of $l$ and $L$ and agrees with $v_{0} = 1.81$meV. 
Agreement between the present one-parameter result for the quadrupole 
correction and the elaborate six-body calculation~\cite{Harston92} 
is a good argument for the validity of the present approach. 

At last, one should obtain $W_{ll_1}^{n}$~(\ref{WN}), which requires 
evaluation 
of $J_{\lambda} \left(1 + \frac{E_{\nu \ell} - E}{\varepsilon_{11}}\right) $ 
by using the $t\mu + d $ scattering phase shifts $\delta_\lambda (k) $ in the 
integrands of Eqs.~(\ref{J0L}) and (\ref{J2L}). The low-energy scattering 
phase shifts were determined in a number of three-body calculations 
\cite{Cohen91,Chiccoli92,Kino93,Igarashi94,Kvitsinsky96,Abramov01}, 
whose results are in good agreement with each other. Using $\delta_\lambda(k)$ 
from these calculations and integrating~(\ref{J0L}) and (\ref{J2L}) 
in the energy interval $0 \le k^2/2\mu_1 \le 10$eV, one obtains 
$J_{\lambda} \left(1 + \frac{E_{\nu \ell} - E}{\varepsilon_{11}}\right) $ 
with a relative accuracy about $0.01$. 

Calculating the matrix elements $V_{ll_1}$~(\ref{Vll}) and 
$W_{ll_1}^{n}$~(\ref{WN}) and solving the eigenvalue 
equation~(\ref{matrixeq}) one obtains energy shifts presented 
in Table~\ref{tabshifts} for $(dt\mu)dee$ and $(dt\mu)tee$. 
\begin{table}[hbt]
\begin{tabular}{r r r r r r}
\multicolumn{2}{c}{} & \multicolumn{2}{c}{$(dt\mu)dee$} &
\multicolumn{2}{c}{$(dt\mu)tee$} \\ \hline
$l$ &\phantom{a} $L$ & \phantom{a} $n = 2$ & $n = 3$ &
\phantom{aa} $n = 2$ & $n = 3$ \\
\hline
$0$ & $1$ & $1.48$ & $0.84$ & $1.70$  & $1.15$ \\

$1$ & $0$ & $1.99$ & $0.55$ & $2.48$  & $1.25$ \\
$1$ & $1$ & $1.22$ & $0.98$ & $1.30$  & $1.09$ \\
$1$ & $2$ & $1.54$ & $0.82$ & $1.78$  & $1.16$ \\

$2$ & $1$ & $1.73$ & $0.66$ & $2.07$  & $1.19$ \\
$2$ & $2$ & $1.22$ & $0.98$ & $1.30$  & $1.09$ \\
$2$ & $3$ & $1.57$ & $0.82$ & $1.82$  & $1.17$ \\

$3$ & $2$ & $1.65$ & $0.68$ & $1.99$  & $1.16$ \\
$3$ & $3$ & $1.22$ & $0.98$ & $1.30$  & $1.09$ \\
$3$ & $4$ & $1.59$ & $0.82$ & $1.84$  & $1.18$ \\

$4$ & $3$ & $1.62$ & $0.68$ & $1.95$  & $1.15$ \\
$4$ & $4$ & $1.22$ & $0.98$ & $1.30$  & $1.09$ \\
$4$ & $5$ & $1.60$ & $0.82$ & $1.86$  & $1.18$ \\
\hline
\end{tabular}
\caption{Energy shifts (meV) for a few states of $(dt\mu)dee$ and 
$(dt\mu)tee$ with the vibrational quantum number $n$, 
the total angular momentum $L$, and angular momentum 
$l$ of the hydrogen-like molecule with the point-like $dt\mu$ quasi-nucleus. } 
\label{tabshifts}
\end{table} 
Note that applicability of the harmonic approximation 
for the BO potential was checked by using the modified dipole matrix 
element~(\ref{U1an}) in the calculation, which gives an estimate of the 
unharmonic correction of the order of $5$\% in the energy shifts. Calculations 
reveal that the energy shifts are essentially dependent on the isotopic 
composition and the molecular quantum numbers $n$ and $l$, which is basically 
connected with the cancellation of the first- and second-order PT 
contributions. 
In particular, the energy shifts decrease with increasing $n$ so that 
$\Delta_{\pm }$ become very small or even negative for $n = 4$. 
The reason for this dependence is an increasing in the dipole matrix 
element~(\ref{U100}) with increasing $n$, which, in turn, leads to an 
increasing in the second-order PT contribution. 
The cancellation effect was widely discussed, e.~g., 
in~\cite{Menshikov85,Scrinzi89a,Harston92a}; nevertheless, 
the dependence on the molecular quantum state was beyond the scope 
of those papers where only the atom-like system $(dt\mu)e$ was 
calculated. On the other hand, the calculation~\cite{Harston92} determined  
the $l$-dependence only in the first-order PT. 

\begin{figure}[htb]
\includegraphics[height=.3\textheight, width=0.45\textwidth]{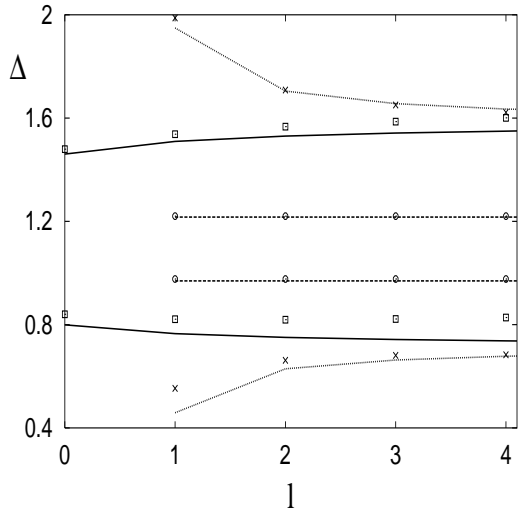} 
\caption{Energy shifts (meV) of $(dt\mu)dee$. Crosses, squares, and circles 
denote, respectively, $\Delta_{+}$, $\Delta_{-}$, and $\Delta_{0}$. 
The results obtained by simplified formulas~(\ref{scalar}), (\ref{enshift}) 
are denoted by the solid, dashed, and dotted lines, respectively. 
Three upper curves correspond to the vibrational quantum number $n = 2$ and 
the lower ones to $n = 3$. 
\label{fig1} }
\end{figure}

The dependence of the energy shifts on quantum numbers 
is illustrated in Figure~\ref{fig1} for the $(dt\mu)dee$ states with 
$n = 2, 3$ and $l = 0 - 4$. 
In addition to a decreasing in the energy shifts for higher $n$, notice the 
inverse ordering of levels, i.~e., the highest level with $L = l - 1$ for 
$n = 2$ becomes the lowest for $n = 3$. Except for $\Delta_{+}$, whose values 
at small $l = 1, 2$ are quite different, the results reveal weak dependence 
on $l$ with splitting of levels of the order of $0.2$meV. 

As discussed at the end of Section~\ref{secshifts}, the dependence of energy 
shifts and level splitting on quantum numbers are expressed to a good accuracy 
by simple formulas~(\ref{scalar}), (\ref{enshift}) via few 
parameters. Numerical values of $\frac{16\omega}{3\pi \varepsilon_{11}}$ and 
$J_{0, 2}^{\pm}$ are given in Table~\ref{tabconstants} for 
$(dt\mu)dee$ and $(dt\mu)tee$. 
\begin{table}[htb]
\begin{tabular}{c@{\extracolsep{1.5mm}} c c c c c}
\multicolumn{6}{c}{}   \\
$X$ & $\frac{16\omega}{3\pi \varepsilon_{11}}$ &
$J_{0}^{+}$ & $J_{0}^{-} $ & $J_{2}^{+}$ & $J_{2}^{-}$  \\
\hline
$d$ & $0.917$ & $0.113$ & $0.236$ & $0.051$ & $0.073$  \\
$t$ & $0.778$ & $0.117$ & $0.216$ & $0.052$ & $0.069$  \\
\hline
\end{tabular}
\caption{Dimensionless parameters entering into the simplified 
expressions~(\ref{scalar}), (\ref{enshift}) for the energy shifts 
of $(dt\mu)Xee$. }
\label{tabconstants}
\end{table}
As is clearly seen in Figure~\ref{fig1}, 
the simplified expressions~(\ref{scalar}), (\ref{enshift}) provide 
a reliable description of energy shifts. 

\section{Discussion and conclusions}

The hydrogen-like molecule $(dt\mu )Xee$ is treated within the framework 
of the three-body model for heavy particles $d$, $t\mu$, and $X$. 
The model is based on the fact that a $t\mu$ mesic atom is small 
in comparison with its mean separation from a deuteron and that
the size of a $dt\mu$ mesic molecule is small in comparison with the 
amplitude of vibrations in $(dt\mu)Xee$. 
In this approach, the interaction of the charged particles $d$ and $X$  
is described by the well-known BO potential of the hydrogen molecule, 
while the description of a $dt\mu $ mesic molecule is given in terms 
of the binding energy $\varepsilon_{11}$, the asymptotic constant $C_a$, 
and the low-energy $s$- and $d$-wave scattering phase shifts 
$\delta_\lambda (k)$ regardless 
of the explicit form of the $d + t\mu $ effective potential. 

In the present approach, the shift and splitting of 
the $(dt\mu)Xee$ energy levels which result from the internal structure 
and motion of the $dt\mu$ quasi-nucleus are calculated in the second-order PT. 
This allows one to find the energy levels, i.~e., the positions of the 
$t\mu + {\mathrm D}{\mathrm X}$ scattering resonances with an accuracy 
about a tenth of a meV, which is of key importance for determination of the 
$dt\mu$ formation rate. Calculations are performed for different vibrational 
($n=2,3$) and rotational ($0\le l \le 4$) states for the molecules of the 
different isotope composition $X=d,t$. In this respect, note that different 
vibrational states of $(dt\mu)Xee$ can be currently observed in the
atomic beam experiments~\cite{Fujiwara00}. It should be emphasized that the 
effect of the $dt\mu$ structure removes the degeneracy of unperturbed states
with the same $l$ and different $L$, which produces a triple-resonance 
structure in place of every unperturbed level except the one with $l = 0$. 
As the splitting value is of the order of the shift itself, the effect 
of splitting should be taken into account in the energy dependence 
of the resonance formation rate. 

The following aspects of the present calculation are worth mentioning. The 
first-order PT quadrupole contribution to the energy shifts is in agreement 
with the elaborate six-body calculation~\cite{Harston92}, which is a good 
argument for the validity of the present approach. Furthermore, it is shown 
that for all the considered states the effect of coupling of the rotational 
states with $l = L - 1$ and $l = L + 1$ is beyond the accuracy of the present 
calculation. In addition, the energy shift and splitting is conveniently 
expressed by simple analytical dependence~(\ref{scalar}), (\ref{enshift}) 
on the rotational-vibrational quantum numbers $n$ and $l$. 

It is of interest to compare the present result with the direct 
non-perturbative three-body calculation of $t\mu + {\mathrm D}_2$ 
scattering~\cite{Zeman00,Zeman01} in which the effect of the $dt\mu$ structure 
is explicitly taken into account. The positions of narrow resonances 
calculated in this paper correspond to energy levels of $(dt\mu)dee$ 
for $L = 0$ ($l = 1$) and $n = 3, 4$. For two approximations of the 
effective potential between $t\mu $ and $d$ used in~\cite{Zeman00,Zeman01}, 
energy shifts are, respectively, $1$meV and $4$meV for $n = 3$ and $2$meV 
and $5$meV for $n = 4$. The cause of the noticeable difference (about $3$meV) 
is not clear since both potentials allow a good description of the low-energy 
properties of the $t\mu + d $ system. The dependence on the choice 
of the effective potential and a limitation only by $L = 0$ hinders 
a quantitative comparison of the present results and those 
of~\cite{Zeman00,Zeman01}. Qualitatively, the energy shifts obtained 
in Ref.~\cite{Zeman00,Zeman01} exceed the present ones and, contrary to PT 
considerations, the value for $n = 4$ is higher than for $n = 3$. 
This $n$-dependence clearly deserves further investigation. 

Finally, it should be mentioned that the present approach, which reliably 
takes into account the structure of the exotic molecule, is promising for 
wider applications, in particular, for determination of resonance positions 
and formation rates beyond PT by solving the scattering problem. Till now,  
except Refs.~\cite{Zeman00,Zeman01}, the formation rates have been calculated 
only in the first-order PT. In this respect, the result of 
Ref.~\cite{Petrov96} shows that the first-order PT dipole approximation 
is questionable and one should do more refined calculations. 
In addition, it is of interest to apply the present approach to the problem 
of the resonance formation of metastable $dt\mu$ mesic 
molecules~\cite{Wallenius96,Wallenius01} in collisions of excited $t\mu $ 
mesic atoms with ${\mathrm D}_2$ molecules.  

\appendix
\section{Angular integrals}
\label{appendix}

The following angular integrals are necessary to calculate the matrix elements 
\begin{equation}
\label{A}
A_{K}^{L}(l\lambda l_1 \lambda_{1})= \int d\hat{\bm{\rho}} d\hat{\mathbf{r}}
P_{K}(\cos{\theta})
{\cal Y}^{LM^{\ast}}_{l\lambda}(\hat{\bm{\rho}}, \hat{\mathbf{r}})
{\cal Y}^{LM}_{l_1 \lambda_1} (\hat{\bm{\rho}}, \hat{\mathbf{r}}) 
\end{equation}
where $\theta$ is the angle between two unit vectors
 $\hat{\mathbf{r}} = \mathbf{r}/r $ and $\hat{\bm{\rho}} = \bm{\rho}/\rho$,
$P_K(x)$ is the Legendre polynomial, and the bispherical harmonics
are defined as
\begin{equation}
{\cal Y}^{LM}_{l \lambda}(\hat{\bm{\rho}}, \hat{\mathbf{r}}) =
\sum_{m \mu}(lm \lambda \mu | LM) Y_{lm}(\hat{\bm{\rho}})
Y_{\lambda \mu}(\hat{\mathbf{r}}) \ .
\end{equation}
Evaluating the integral~(\ref{A}) one comes to the expression in terms of
the Clebsh-Gordon coefficients and $6j$-symbols
\begin{equation}
\label{A6j}
A_{K}^{L}(l\lambda l_1 \lambda_{1})=(-)^{l_{1}+L}
\sqrt{(2l+1)(2\lambda+1)} (l0K0\mid l_1 0) (\lambda 0K0\mid \lambda_1 0)
\left\{\begin{array}{ccc}l_{1}&l&K\\
\lambda&\lambda_{1}&L\end{array}\right\} \ .
\end{equation}

The matrix elements $V_{ll_{1}}^{n}$ (\ref{matrel1mod}) are expressed in terms 
of the integrals~(\ref{A6j}) with $K=2$ and $\lambda = \lambda_1 = 1$ which 
are explicitly written as 
\begin{equation}
\label{Aquadro}
\begin{array}{c}
A_{2}^{L}(l1l_1 1) = \displaystyle{- \frac{1}{5} \delta_{lL} \delta_{l_1L} + 
\frac{(L+2)}{5(2L+1)}\delta_{lL+1} \delta_{l_1 L+1} +
\frac{(L-1)}{5(2L+1)}\delta_{l L-1}\delta_{l_1 L-1} - }\\
\displaystyle{\frac{3}{5}\frac{\sqrt{L(L+1)}}{(2L+1)} 
(\delta_{l L-1} \delta_{l_1 L+1} + \delta_{l L+1} \delta_{l_1 L-1})} \ .
\end{array}
\end{equation}
The matrix elements $W_{ll_1}^{n} $ (\ref{W}) are expressed in terms of 
the integrals (\ref{A6j}) with $K=1$, $\lambda=1$, and  
either $\lambda_1=0$ and $l_{1} = L$ or $\lambda_1=2$ and $l=L\pm 1$.
The explicit expressions read
\begin{eqnarray}
\label{Ad0}
A_{1}^{L}(l 1 L 0)
& = &\frac{1}{\sqrt{3 (2L+1)}}\left(\sqrt{L} \delta_{l,L-1}-\sqrt{L+1}
\delta_{l, L+1}\right)\\
\label{Ad2}
A_1^L(l1l_12)& = &\left\{
\begin{array}{c}
-\displaystyle\frac{1}{2}\sqrt{\frac{(L+l+3)(L+l+4)(L-l-2)(L-l-3)}{
  15(2l+1)(2l+3)}},\qquad l_1 = l+1\\
\displaystyle\frac{1}{2}\sqrt{\frac{(L-l+2)(L-l+3)(L+l-2)(L+l-1)}{
  15(4l^2-1)}},\qquad l_1 = l-1
\end{array}
\right.
\end{eqnarray}

\bibliography{dtmu}

\begin{thebibliography}{28}
\expandafter\ifx\csname natexlab\endcsname\relax\def\natexlab#1{#1}\fi
\expandafter\ifx\csname bibnamefont\endcsname\relax
  \def\bibnamefont#1{#1}\fi
\expandafter\ifx\csname bibfnamefont\endcsname\relax
  \def\bibfnamefont#1{#1}\fi
\expandafter\ifx\csname citenamefont\endcsname\relax
  \def\citenamefont#1{#1}\fi
\expandafter\ifx\csname url\endcsname\relax
  \def\url#1{\texttt{#1}}\fi
\expandafter\ifx\csname urlprefix\endcsname\relax\def\urlprefix{URL }\fi
\providecommand{\bibinfo}[2]{#2}
\providecommand{\eprint}[2][]{\url{#2}}

\bibitem[{\citenamefont{Breunlich et~al.}(1989)\citenamefont{Breunlich, Kammel,
  Cohen, and Leon}}]{Breunlich89}
\bibinfo{author}{\bibfnamefont{W.~H.} \bibnamefont{Breunlich}},
  \bibinfo{author}{\bibfnamefont{P.}~\bibnamefont{Kammel}},
  \bibinfo{author}{\bibfnamefont{J.~S.} \bibnamefont{Cohen}}, \bibnamefont{and}
  \bibinfo{author}{\bibfnamefont{M.}~\bibnamefont{Leon}},
  \bibinfo{journal}{Annu.\ Rev.\ Nucl.\ Part.\ Sci.}
  \textbf{\bibinfo{volume}{39}}, \bibinfo{pages}{311} (\bibinfo{year}{1989}).

\bibitem[{\citenamefont{Ponomarev}(1990)}]{Ponomarev90}
\bibinfo{author}{\bibfnamefont{L.~I.} \bibnamefont{Ponomarev}},
  \bibinfo{journal}{Contemp.\ Phys.} \textbf{\bibinfo{volume}{31}},
  \bibinfo{pages}{219} (\bibinfo{year}{1990}).

\bibitem[{\citenamefont{Rafelski et~al.}(1991)\citenamefont{Rafelski, Harley,
  Shin, and Rafelski}}]{Rafelski91}
\bibinfo{author}{\bibfnamefont{H.~E.} \bibnamefont{Rafelski}},
  \bibinfo{author}{\bibfnamefont{D.}~\bibnamefont{Harley}},
  \bibinfo{author}{\bibfnamefont{G.~R.} \bibnamefont{Shin}}, \bibnamefont{and}
  \bibinfo{author}{\bibfnamefont{J.}~\bibnamefont{Rafelski}},
  \bibinfo{journal}{J.\ Phys. B} \textbf{\bibinfo{volume}{24}},
  \bibinfo{pages}{1469} (\bibinfo{year}{1991}).

\bibitem[{\citenamefont{Froelich}(1992)}]{Froelich92}
\bibinfo{author}{\bibfnamefont{P.}~\bibnamefont{Froelich}},
  \bibinfo{journal}{Adv.\ Phys.} \textbf{\bibinfo{volume}{41}},
  \bibinfo{pages}{405} (\bibinfo{year}{1992}).

\bibitem[{\citenamefont{Vesman}(1967)}]{Vesman67}
\bibinfo{author}{\bibfnamefont{E.~A.} \bibnamefont{Vesman}},
  \bibinfo{journal}{Pis'ma Zh.\ Eksp.\ Teor.\ Fiz}
  \textbf{\bibinfo{volume}{5}}, \bibinfo{pages}{113} (\bibinfo{year}{1967}),
  \bibinfo{note}{[JETP Lett. {\textbf 5}, 91 (1967)]}.

\bibitem[{\citenamefont{Faifman et~al.}(1986)\citenamefont{Faifman, Menshikov,
  Ponomarev, Puzynin, Puzynina, and Strizh}}]{Faifman86}
\bibinfo{author}{\bibfnamefont{M.~P.} \bibnamefont{Faifman}},
  \bibinfo{author}{\bibfnamefont{L.~I.} \bibnamefont{Menshikov}},
  \bibinfo{author}{\bibfnamefont{L.~I.} \bibnamefont{Ponomarev}},
  \bibinfo{author}{\bibfnamefont{I.~V.} \bibnamefont{Puzynin}},
  \bibinfo{author}{\bibfnamefont{T.~P.} \bibnamefont{Puzynina}},
  \bibnamefont{and} \bibinfo{author}{\bibfnamefont{T.~A.}
  \bibnamefont{Strizh}}, \bibinfo{journal}{Z.\ Phys. A}
  \textbf{\bibinfo{volume}{2}}, \bibinfo{pages}{79} (\bibinfo{year}{1986}).

\bibitem[{\citenamefont{Scrinzi et~al.}(1988)\citenamefont{Scrinzi, Szalewicz,
  and Monkhorst}}]{Scrinzi88}
\bibinfo{author}{\bibfnamefont{A.}~\bibnamefont{Scrinzi}},
  \bibinfo{author}{\bibfnamefont{K.}~\bibnamefont{Szalewicz}},
  \bibnamefont{and} \bibinfo{author}{\bibfnamefont{H.~J.}
  \bibnamefont{Monkhorst}}, \bibinfo{journal}{Phys.\ Rev. A}
  \textbf{\bibinfo{volume}{37}}, \bibinfo{pages}{2270} (\bibinfo{year}{1988}).

\bibitem[{\citenamefont{Harston
  et~al.}(1992{\natexlab{a}})\citenamefont{Harston, Shimamura, and
  Kamimura}}]{Harston92}
\bibinfo{author}{\bibfnamefont{M.~R.} \bibnamefont{Harston}},
  \bibinfo{author}{\bibfnamefont{I.}~\bibnamefont{Shimamura}},
  \bibnamefont{and} \bibinfo{author}{\bibfnamefont{M.}~\bibnamefont{Kamimura}},
  \bibinfo{journal}{Phys.\ Rev. A} \textbf{\bibinfo{volume}{45}},
  \bibinfo{pages}{94} (\bibinfo{year}{1992}{\natexlab{a}}).

\bibitem[{\citenamefont{Menshikov}(1985)}]{Menshikov85}
\bibinfo{author}{\bibfnamefont{L.~I.} \bibnamefont{Menshikov}},
  \bibinfo{journal}{Yad.\ Fiz.} \textbf{\bibinfo{volume}{42}},
  \bibinfo{pages}{1449} (\bibinfo{year}{1985}).

\bibitem[{\citenamefont{Scrinzi and Szalewicz}(1989)}]{Scrinzi89a}
\bibinfo{author}{\bibfnamefont{A.}~\bibnamefont{Scrinzi}} \bibnamefont{and}
  \bibinfo{author}{\bibfnamefont{K.}~\bibnamefont{Szalewicz}},
  \bibinfo{journal}{Phys.\ Rev. A} \textbf{\bibinfo{volume}{39}},
  \bibinfo{pages}{4983} (\bibinfo{year}{1989}).

\bibitem[{\citenamefont{Zeman et~al.}(2000)\citenamefont{Zeman, Armour, and
  Pack}}]{Zeman00}
\bibinfo{author}{\bibfnamefont{V.}~\bibnamefont{Zeman}},
  \bibinfo{author}{\bibfnamefont{E.~A.~G.} \bibnamefont{Armour}},
  \bibnamefont{and} \bibinfo{author}{\bibfnamefont{R.~T.} \bibnamefont{Pack}},
  \bibinfo{journal}{Phys.\ Rev. A} \textbf{\bibinfo{volume}{61}},
  \bibinfo{pages}{052713} (\bibinfo{year}{2000}).

\bibitem[{\citenamefont{Zeman and Armour}(2001)}]{Zeman01}
\bibinfo{author}{\bibfnamefont{V.}~\bibnamefont{Zeman}} \bibnamefont{and}
  \bibinfo{author}{\bibfnamefont{E.~A.~G.} \bibnamefont{Armour}},
  \bibinfo{journal}{Hyperfine \ Interact.} \textbf{\bibinfo{volume}{138}},
  \bibinfo{pages}{255} (\bibinfo{year}{2001}).

\bibitem[{\citenamefont{Sharp}(1971)}]{Sharp71}
\bibinfo{author}{\bibfnamefont{T.~E.} \bibnamefont{Sharp}},
  \bibinfo{journal}{Atomic \ Data} \textbf{\bibinfo{volume}{2}},
  \bibinfo{pages}{119} (\bibinfo{year}{1971}).

\bibitem[{\citenamefont{Kolos and Wolniewicz}(1964)}]{Kolos64}
\bibinfo{author}{\bibfnamefont{W.}~\bibnamefont{Kolos}} \bibnamefont{and}
  \bibinfo{author}{\bibfnamefont{L.}~\bibnamefont{Wolniewicz}},
  \bibinfo{journal}{J.\ Chem.\ Phys.} \textbf{\bibinfo{volume}{41}},
  \bibinfo{pages}{3663} (\bibinfo{year}{1964}).

\bibitem[{\citenamefont{Kolos et~al.}(1986)\citenamefont{Kolos, Szalewicz, and
  Monkhorst}}]{Kolos86}
\bibinfo{author}{\bibfnamefont{W.}~\bibnamefont{Kolos}},
  \bibinfo{author}{\bibfnamefont{K.}~\bibnamefont{Szalewicz}},
  \bibnamefont{and} \bibinfo{author}{\bibfnamefont{H.~J.}
  \bibnamefont{Monkhorst}}, \bibinfo{journal}{J.\ Chem.\ Phys.}
  \textbf{\bibinfo{volume}{84}}, \bibinfo{pages}{3278} (\bibinfo{year}{1986}).

\bibitem[{\citenamefont{Aissing et~al.}(1990)\citenamefont{Aissing, Monkhorst,
  and Petrov}}]{Aissing90}
\bibinfo{author}{\bibfnamefont{G.}~\bibnamefont{Aissing}},
  \bibinfo{author}{\bibfnamefont{H.~J.} \bibnamefont{Monkhorst}},
  \bibnamefont{and} \bibinfo{author}{\bibfnamefont{Y.~V.}
  \bibnamefont{Petrov}}, \bibinfo{journal}{Phys.\ Rev. A}
  \textbf{\bibinfo{volume}{42}}, \bibinfo{pages}{6894} (\bibinfo{year}{1990}).

\bibitem[{\citenamefont{Kino et~al.}(1995)\citenamefont{Kino, Harston,
  Shimamura, Armour, and Kamimura}}]{Kino95}
\bibinfo{author}{\bibfnamefont{Y.}~\bibnamefont{Kino}},
  \bibinfo{author}{\bibfnamefont{M.~R.} \bibnamefont{Harston}},
  \bibinfo{author}{\bibfnamefont{I.}~\bibnamefont{Shimamura}},
  \bibinfo{author}{\bibfnamefont{E.~A.~G.} \bibnamefont{Armour}},
  \bibnamefont{and} \bibinfo{author}{\bibfnamefont{M.}~\bibnamefont{Kamimura}},
  \bibinfo{journal}{Phys.\ Rev. A} \textbf{\bibinfo{volume}{52}},
  \bibinfo{pages}{870} (\bibinfo{year}{1995}).

\bibitem[{\citenamefont{Harston
  et~al.}(1992{\natexlab{b}})\citenamefont{Harston, Shimamura, and
  Kamimura}}]{Harston92a}
\bibinfo{author}{\bibfnamefont{M.~R.} \bibnamefont{Harston}},
  \bibinfo{author}{\bibfnamefont{I.}~\bibnamefont{Shimamura}},
  \bibnamefont{and} \bibinfo{author}{\bibfnamefont{M.}~\bibnamefont{Kamimura}},
  \bibinfo{journal}{Z.\ Phys. D} \textbf{\bibinfo{volume}{22}},
  \bibinfo{pages}{635} (\bibinfo{year}{1992}{\natexlab{b}}).

\bibitem[{\citenamefont{Cohen and Struensee}(1991)}]{Cohen91}
\bibinfo{author}{\bibfnamefont{J.~S.} \bibnamefont{Cohen}} \bibnamefont{and}
  \bibinfo{author}{\bibfnamefont{M.}~\bibnamefont{Struensee}},
  \bibinfo{journal}{Phys.\ Rev. A} \textbf{\bibinfo{volume}{43}},
  \bibinfo{pages}{3460} (\bibinfo{year}{1991}).

\bibitem[{\citenamefont{Chiccoli et~al.}(1992)\citenamefont{Chiccoli, Korobov,
  Melezhik, Pasini, Ponomarev, and Wozniak}}]{Chiccoli92}
\bibinfo{author}{\bibfnamefont{C.}~\bibnamefont{Chiccoli}},
  \bibinfo{author}{\bibfnamefont{V.~I.} \bibnamefont{Korobov}},
  \bibinfo{author}{\bibfnamefont{V.~S.} \bibnamefont{Melezhik}},
  \bibinfo{author}{\bibfnamefont{P.}~\bibnamefont{Pasini}},
  \bibinfo{author}{\bibfnamefont{L.~I.} \bibnamefont{Ponomarev}},
  \bibnamefont{and} \bibinfo{author}{\bibfnamefont{J.}~\bibnamefont{Wozniak}},
  \bibinfo{journal}{Muon \ Catal.\ Fusion} \textbf{\bibinfo{volume}{7}},
  \bibinfo{pages}{87} (\bibinfo{year}{1992}).

\bibitem[{\citenamefont{Kino and Kamimura}(1993)}]{Kino93}
\bibinfo{author}{\bibfnamefont{Y.}~\bibnamefont{Kino}} \bibnamefont{and}
  \bibinfo{author}{\bibfnamefont{M.}~\bibnamefont{Kamimura}},
  \bibinfo{journal}{Hyperfine \ Interact.} \textbf{\bibinfo{volume}{82}},
  \bibinfo{pages}{45} (\bibinfo{year}{1993}).

\bibitem[{\citenamefont{Igarashi et~al.}(1994)\citenamefont{Igarashi, Toshima,
  and Shirai}}]{Igarashi94}
\bibinfo{author}{\bibfnamefont{A.}~\bibnamefont{Igarashi}},
  \bibinfo{author}{\bibfnamefont{N.}~\bibnamefont{Toshima}}, \bibnamefont{and}
  \bibinfo{author}{\bibfnamefont{T.}~\bibnamefont{Shirai}},
  \bibinfo{journal}{Phys.\ Rev. A} \textbf{\bibinfo{volume}{50}},
  \bibinfo{pages}{4951} (\bibinfo{year}{1994}).

\bibitem[{\citenamefont{Kvitsinsky et~al.}(1996)\citenamefont{Kvitsinsky, Hu,
  and Cohen}}]{Kvitsinsky96}
\bibinfo{author}{\bibfnamefont{A.~A.} \bibnamefont{Kvitsinsky}},
  \bibinfo{author}{\bibfnamefont{C.-Y.} \bibnamefont{Hu}}, \bibnamefont{and}
  \bibinfo{author}{\bibfnamefont{J.~S.} \bibnamefont{Cohen}},
  \bibinfo{journal}{Phys.\ Rev. A} \textbf{\bibinfo{volume}{53}},
  \bibinfo{pages}{255} (\bibinfo{year}{1996}).

\bibitem[{\citenamefont{Abramov et~al.}(2001)\citenamefont{Abramov, Gusev, and
  Ponomarev}}]{Abramov01}
\bibinfo{author}{\bibfnamefont{D.~I.} \bibnamefont{Abramov}},
  \bibinfo{author}{\bibfnamefont{V.~V.} \bibnamefont{Gusev}}, \bibnamefont{and}
  \bibinfo{author}{\bibfnamefont{L.~I.} \bibnamefont{Ponomarev}},
  \bibinfo{journal}{Yad. \ Fiz.} \textbf{\bibinfo{volume}{64}},
  \bibinfo{pages}{1442} (\bibinfo{year}{2001}).

\bibitem[{\citenamefont{Fujiwara et~al.}(2000)\citenamefont{Fujiwara, Adamczak,
  Bailey, Beerand, Beveridgeand, Faifman, Huberand, Kammel, Kim, Knowles
  et~al.}}]{Fujiwara00}
\bibinfo{author}{\bibfnamefont{M.~C.} \bibnamefont{Fujiwara}},
  \bibinfo{author}{\bibfnamefont{A.}~\bibnamefont{Adamczak}},
  \bibinfo{author}{\bibfnamefont{J.~M.} \bibnamefont{Bailey}},
  \bibinfo{author}{\bibfnamefont{G.~A.} \bibnamefont{Beerand}},
  \bibinfo{author}{\bibfnamefont{J.~L.} \bibnamefont{Beveridgeand}},
  \bibinfo{author}{\bibfnamefont{M.~P.} \bibnamefont{Faifman}},
  \bibinfo{author}{\bibfnamefont{T.~M.} \bibnamefont{Huberand}},
  \bibinfo{author}{\bibfnamefont{P.}~\bibnamefont{Kammel}},
  \bibinfo{author}{\bibfnamefont{S.~K.} \bibnamefont{Kim}},
  \bibinfo{author}{\bibfnamefont{P.~E.} \bibnamefont{Knowles}},
  \bibnamefont{et~al.}, \bibinfo{journal}{Phys.\ Rev.\ Lett.}
  \textbf{\bibinfo{volume}{85}}, \bibinfo{pages}{1642} (\bibinfo{year}{2000}).

\bibitem[{\citenamefont{Petrov and Petrov}(1996)}]{Petrov96}
\bibinfo{author}{\bibfnamefont{Y.~V.} \bibnamefont{Petrov}} \bibnamefont{and}
  \bibinfo{author}{\bibfnamefont{V.~Y.} \bibnamefont{Petrov}},
  \bibinfo{journal}{Phys.\ Lett. B} \textbf{\bibinfo{volume}{378}},
  \bibinfo{pages}{1} (\bibinfo{year}{1996}).

\bibitem[{\citenamefont{Wallenius and Froelich}(1996)}]{Wallenius96}
\bibinfo{author}{\bibfnamefont{J.}~\bibnamefont{Wallenius}} \bibnamefont{and}
  \bibinfo{author}{\bibfnamefont{P.}~\bibnamefont{Froelich}},
  \bibinfo{journal}{Phys.\ Rev. A} \textbf{\bibinfo{volume}{54}},
  \bibinfo{pages}{1171} (\bibinfo{year}{1996}).

\bibitem[{\citenamefont{Wallenius et~al.}(2001)\citenamefont{Wallenius,
  Jonsell, Kino, and Froelich}}]{Wallenius01}
\bibinfo{author}{\bibfnamefont{J.}~\bibnamefont{Wallenius}},
  \bibinfo{author}{\bibfnamefont{S.}~\bibnamefont{Jonsell}},
  \bibinfo{author}{\bibfnamefont{Y.}~\bibnamefont{Kino}}, \bibnamefont{and}
  \bibinfo{author}{\bibfnamefont{P.}~\bibnamefont{Froelich}},
  \bibinfo{journal}{Hyperfine \ Interact.} \textbf{\bibinfo{volume}{138}},
  \bibinfo{pages}{285} (\bibinfo{year}{2001}).

\end{thebibliography}

\end{document}